\documentclass[aps,pra,amssymb,twocolumn,superscriptaddress,showpacs]{revtex4-1}

\usepackage[pdftex]{graphicx}
\usepackage{dcolumn}
\usepackage{bm}
\usepackage{color}

\begin{document}

\title{Further evidence of antibunching of two coherent beams of fermions}
\author{M. Iannuzzi}
\affiliation{Dipartimento di Fisica, Universit\`a di Roma ``Tor Vergata", I-00133 Roma, Italy}
\author{R. Messi}
\affiliation{Dipartimento di Fisica, Universit\`a di Roma ``Tor Vergata", I-00133 Roma, Italy}
\affiliation{INFN Sezione ``Roma Tor Vergata", I-00133 Roma, Italy}
\author{D. Moricciani}
\affiliation{INFN Sezione ``Roma Tor Vergata", I-00133 Roma, Italy}
\author{A. Orecchini}
\affiliation{Dipartimento di Fisica, Universit\`{a} di Perugia, I-06123 Perugia, Italy} 
\author{F. Sacchetti}
\affiliation{Dipartimento di Fisica, Universit\`{a} di Perugia, I-06123 Perugia, Italy} 
\affiliation{CNR-INFM, Centro di Ricerca e Sviluppo SOFT, I-00185 Roma, Italy}
\author{P. Facchi}
\affiliation{Dipartimento di Matematica and MECENAS, Universit\`a di Bari, I-70125  Bari, Italy}
\affiliation{INFN, Sezione di Bari, I-70126 Bari, Italy}
\author{S. Pascazio} 
\affiliation{Dipartimento di Fisica and MECENAS, Universit\`a di Bari, I-70126  Bari, Italy}
\affiliation{INFN, Sezione di Bari, I-70126 Bari, Italy}

\date{\today}

\begin{abstract}
We describe an experiment confirming the evidence of the antibunching effect on a beam of non interacting thermal neutrons. The comparison between the results recorded with a high energy-resolution source of neutrons and those recorded with a broad 
energy-resolution source enables us to clarify the role played by the beam coherence in the occurrence of the 
antibunching effect.  
\end{abstract}

\pacs{03.75.-b; 03.75.Dg}

\maketitle


Quantum correlations have been observed on many fundamental particles, proving their quantum nature. The first observation was done on photons in astronomical research and is a cornerstone in fundamental physics, known as Hanbury Brown-Twiss effect \cite{bib2}. Since then, many experiments have been performed, both on bosonic and fermionic systems, such as electrons, pions and atoms \cite{bib4,bib8,bib5,bib6,CERN,bib7,atom,bib3}. 

The crucial difference between the Bose-Einstein and Fermi-Dirac statistics are the phase space densities, that can change by several orders of magnitude. In a laser
beam, the density is of order $10^{14}$, while typical densities
for thermal light, synchrotron radiation and electrons are of order
$10^{-3}$; finally, for the most advanced neutron sources, one gets
$10^{-15}$. Since bunching and antibunching are 
second-order coherence effects \cite{sudarshan,glauber,bargmann,Loudon,MandelWolf}, 
the above figures make it very difficult to observe fermion
antibunching \cite{bib13}. 

In a previous Letter \cite{bib1} we reported on a coincidence experiment performed on a beam of free noninteracting thermal neutrons. The experiment was a massive-particle analogue of the seminal optical Hanbury-Brown and Twiss experiment on photon bunching \cite{bib2}, and its outcome proved the occurrence of antibunching on pairs of free fermions in the triplet spin state. Although other experiments evidenced antibunching on various fermionic systems, like electrons and atoms \cite{bib3,bib4,bib5,bib6,bib7,atom,bib8}, a basic question remained controversial in relation to the free neutron experiment \cite{bib1}. In order to observe antibunching in a coincidence experiment, a prerequisite is that the active surfaces of the two detectors, as viewed from the source, belong to the same {\it transverse} coherence area of the fermion beam and are shined by a reasonable number of particle pairs from such area. Now, it may be argued that, even for beams of thermal neutrons produced by the most advanced sources, such a condition cannot be realistically satisfied, because by assuming a thermal neutron flux of the order of 10$^{15}\,$n~cm$^{-2}$s$^{-1}$, the mean number of neutrons within the phase space  volume is extremely low and normally smaller than 10$^{-14}$. This problem has been recently discussed in detail and no simple explanation for the experimental observation has been proposed \cite{bib9}. From the theoretical discussion of Ref.\ \cite{bib9}
it emerges that in the actual experiment the preparation of the incoming beam is such that a lateral correlation of the neutron pairs could be introduced by the high resolution monochromator which is employed to obtain an adequate longitudinal coherence.
 
In order to tackle this question, we performed a new twofold-purpose experiment, trying also to maximize the statistics. A first part of the new experiment is based on an advanced acquisition system and employs the same highly monochromatic source ($\Delta E < 0.1~\mu$eV) used in the previous experiment \cite{bib1}. The second part of the experiment is identical to the first one, apart from the use of a source that provides a less monochromatic neutron beam ($\Delta E \simeq 1$meV), so that one has a negligible {\it longitudinal} coherence. In this paper we present the results of this new experiment, that confirm the previous observation \cite{bib1} also by comparison with the low-coherence part of the test.
  
\begin{figure}
\begin{center}
\includegraphics[width=0.9\linewidth]{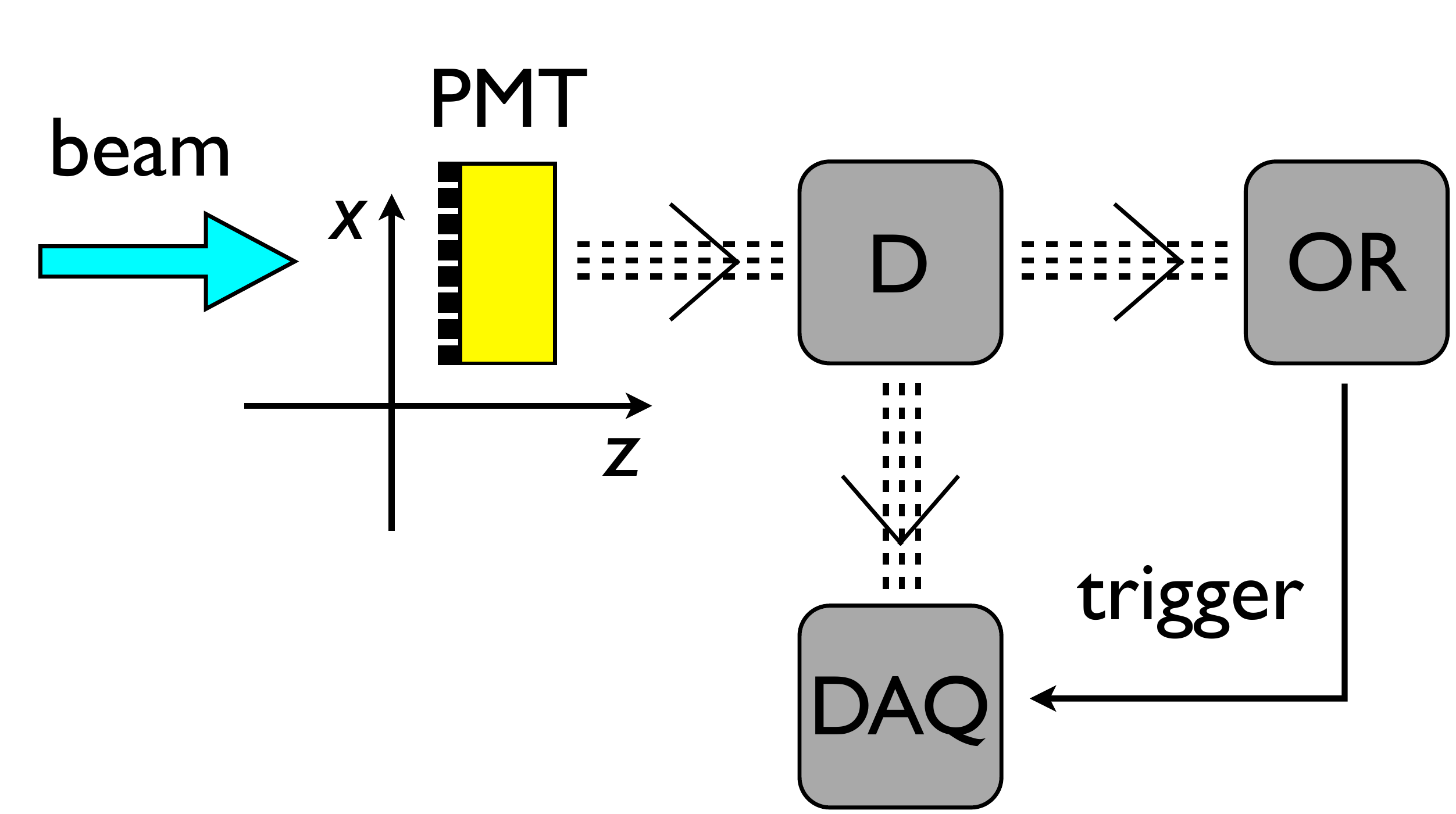}
\end{center}
\vspace{-0.5cm}
\caption{Schematic drawing of the experimental setup. PMT: $8\times8$ multi-anode Hamamatsu photomultiplier (pixel size: $5.8\,$mm$\,\times\, 5.8\,$mm). The 64 signals from the PMT are discriminated by Le Croy 4413 modules D that provide two outputs for each channel. DAQ: data acquisition system. OR: logical circuit that triggers DAQ if D forwards any signal from PMT.
}
\label{fig1}
\end{figure}  
  
As in the previous experiment \cite{bib1}, the present measurements have been performed at the Institute Laue Langevin (Grenoble, France) which provides high quality neutron beam lines. The first part of the experiment was performed by using the primary spectrometer of the IN10 beam line, which produces a monochromatic beam of thermal neutrons from an almost perfect Si(111) single crystal in the perfect backscattering configuration. The small size of the monochromating crystal ($10\times10\,$cm$^2$) and the distance from the detector ($\simeq 10\,$m) guarantee, at an incoming neutron energy of $2.08\,$meV, a very sharp energy window, which can be estimated from the monochromator geometry to be $\Delta E \leq 0.02\,\mu$eV, by assuming a perfect alignment of the monochromating crystal. Unfortunately, the small size of the monochromator and the sharp energy window provide a rather small total rate of $\Phi  \simeq 3000 \,$ n s$^{-1}$ on a transverse size of $\simeq 4.5 \times 3.5\,$cm$^2$.

The second part of the experiment was performed by using the primary spectrometer of the test beam line T13C where a relatively low angle monochromator is used. In this way the energy window is fairly broad, of the order of 0.2~meV at an incoming energy of about 25~meV. In such a situation the coherence time is so short that the antibunching effect cannot be detectable, allowing these results to be used as a reference. To compare the two sets of data on the same ground the detector and the acquisition system of the first experiment has been employed without any change.
      
The coincidence experiment, whose scheme is represented in Fig.\ \ref{fig1}, was performed by using a position-sensitive detector and an advanced electronic acquisition system for the coincidence recording. The detector was a Hamamatsu H8500 multi-anode photomultiplier (PMT) with $8\times8$ pixels (pixel size $5.8\,$mm$\times 5.8\,$mm) coupled to a 0.20 mm thick scintillator ($^6$Li 98\% enriched lithium glass GS20, Applied Scintillation Technologies) directly coupled to the anode window by using an optical grease. The multi-anode configuration of this detector allowed us to arrange the anode pixels into two distinct groupings, e.g.\ vertical or horizontal, forming two detectors separated in the $x$-$y$ plane perpendicular to the neutron beam and at the same position in the $z$ direction along the neutron beam direction. The signals from such two detectors could be time-correlated by detecting their coincidence rate $c(t)$ as a function of their relative time delay $t$, that is, as a function of their virtual separation along the longitudinal direction. The signal produced on each anode is first discriminated against the low amplitude noise produced by the PMT and by the background $\gamma$-radiation and then transmitted to the acquisition system. The present acquisition system is designed for detecting the neutron coincidences with a good time resolution. To this purpose all neutron arrival times on each pixel are recorded in a file for an off line data analysis. All the arrival times were measured by using an internal 40~MHz clock which provides 25~ns time resolution, shorter than the light decay time of the lithium glass scintillator (250~ns) and the maximum traveling time through the scintillator itself ($\simeq$ 300 ns, see discussion below). The data acquisition system (DAQ) recorded events during repeated cycles lasting 10~s each and the transfer process to the computer was performed in about 10 ms, so that the effective duty cycle was about 1000. The whole experiment on IN10 took data over 135 hours.

\begin{figure}
\vspace{-0.9cm}
\begin{center}
\includegraphics[angle=-90,width=1.15\linewidth]{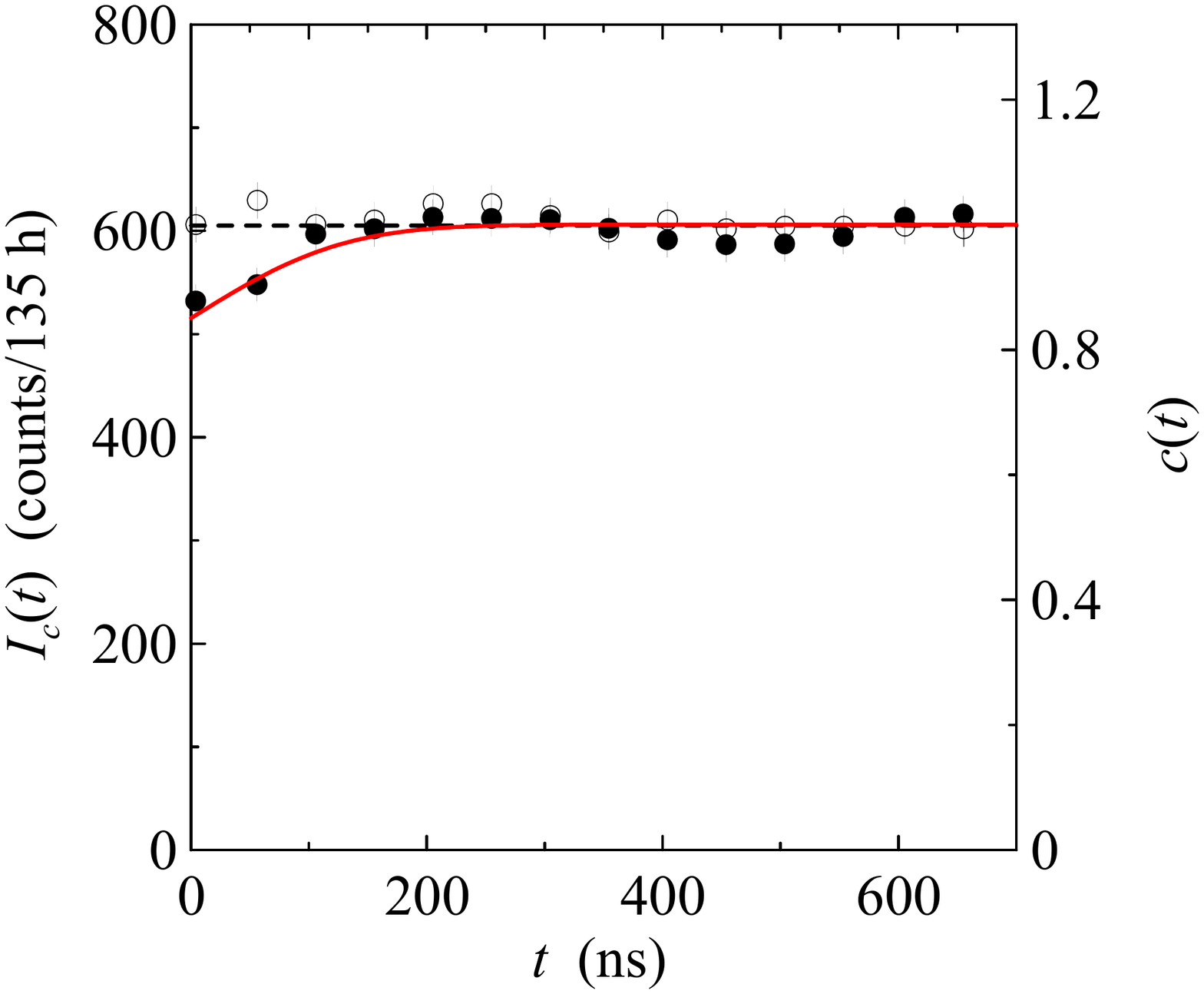}
\end{center}
\vspace{-1.2cm}
\caption{Number of coincidences as a function of the neutron delay time along the traveling direction $z$. Two 6-pixel columns with a relative separation of 3.0~cm; $\Delta = 400\,$ns. 
Full circles: measurements at the high energy resolution source IN10. 
Open circles: measurements at the broad energy resolution source T13C, normalized with respect to the round points around $t = 600\,$ns. Full line: fit from Eq.\ (\ref{eq:cexp}).}
\label{fig2}
\end{figure}

Spurious coincidences are present due to the cross-talk of light emitted in the scintillator as a consequence of neutron absorption, or due to other sources of noise, like the dark counts of the PMT and electronic noise. By analysing the data of our experiment, the estimated time spread of most spurious coincidences, $\delta_s$, was limited to the interval $\delta_s \simeq 150\,$ns. In order to suppress such unwanted coincidences in the data analysis, we have adopted the following procedure, justified by the very low expected probability of having two neutrons within the coherence time $\tau_c$, since $\Phi \tau_c \ll 1$. 
If there were more than one event within a time window $\delta_s$ 
on the first pixel grouping or on the second one, i.e. if two or more pixels of the same pixel grouping lighted up in the time window $\delta_s$, those events were taken as equivalent to one event only. In addition, all events due to lighting up of the first or the second group of pixels which were in the same time window $\delta_s$ with any other pixel of the multi-anode PMT not belonging to these two groups were not considered. This procedure proved to be quite efficient to suppress most of the cross-talk effect. Nevertheless, we expect that some residual cross-talk counts could not be completely eliminated, thus affecting the measurements at small time separations.

The final coincidence detection was based on the determination of the arrival times of the two signals produced respectively at a first pixel grouping used as first detector ($D_1$ at $t_1 =\;$start) and a second pixel grouping, used as second detector ($D_2$ at $t_2 =\;$stop). The pixel grouping can be adjusted to get the best statistics and the best contrast to enhance the presence of neutron antibunching. The coincidences were measured by collecting all the neutrons recorded by $D_2$ at time $t_2>t_1$ ($t_1=$start time) within a time window $\Delta=400\,$ns.
In this way a good statistics is obtained on the coincidences, while no broadening of the correlation function is produced, apart from that due to neutron collection within the scintillator. Since the scintillator has a finite width, as mentioned above, neutrons can be captured within the corresponding traveling time. However, by taking into account the capture probability profile along the depth of the scintillator, it is found that the average capture time from the neutron entrance into the scintillator is about 100 ns. Some broadening effect is also expected from the light decay curve, this effect being also of the order of 100 ns. Therefore, in the present experimental set-up the antibunching effect should be confined within a window larger than some 150 ns. Of course the present counting system reduces the dip expected at small time separation by a factor proportional to $\tau_c / \Delta$. In conclusion, we are measuring an experimental coincidence rate 
\begin{equation}
c_{\mathrm{exp}}(t) = {1 \over \Delta} \, \int_t^{t + \Delta} \, (W * c) (t') \, dt', \qquad t \ge 0,
\end{equation}
where the convolution with $W(t)$ accounts for the broadening of the correlation function due to both the scintillator thickness and the light decay curve. We applied the same analysis process to all data, i.e.\ those measured at the IN10 beam line, where the coherence time $\tau_c$ is rather long, and those measured at the T13C beam line, where $\tau_c$ is much shorter.

\begin{figure}
\vspace{-0.9cm}
\begin{center}
\includegraphics[angle=-90,width=1.15\linewidth]{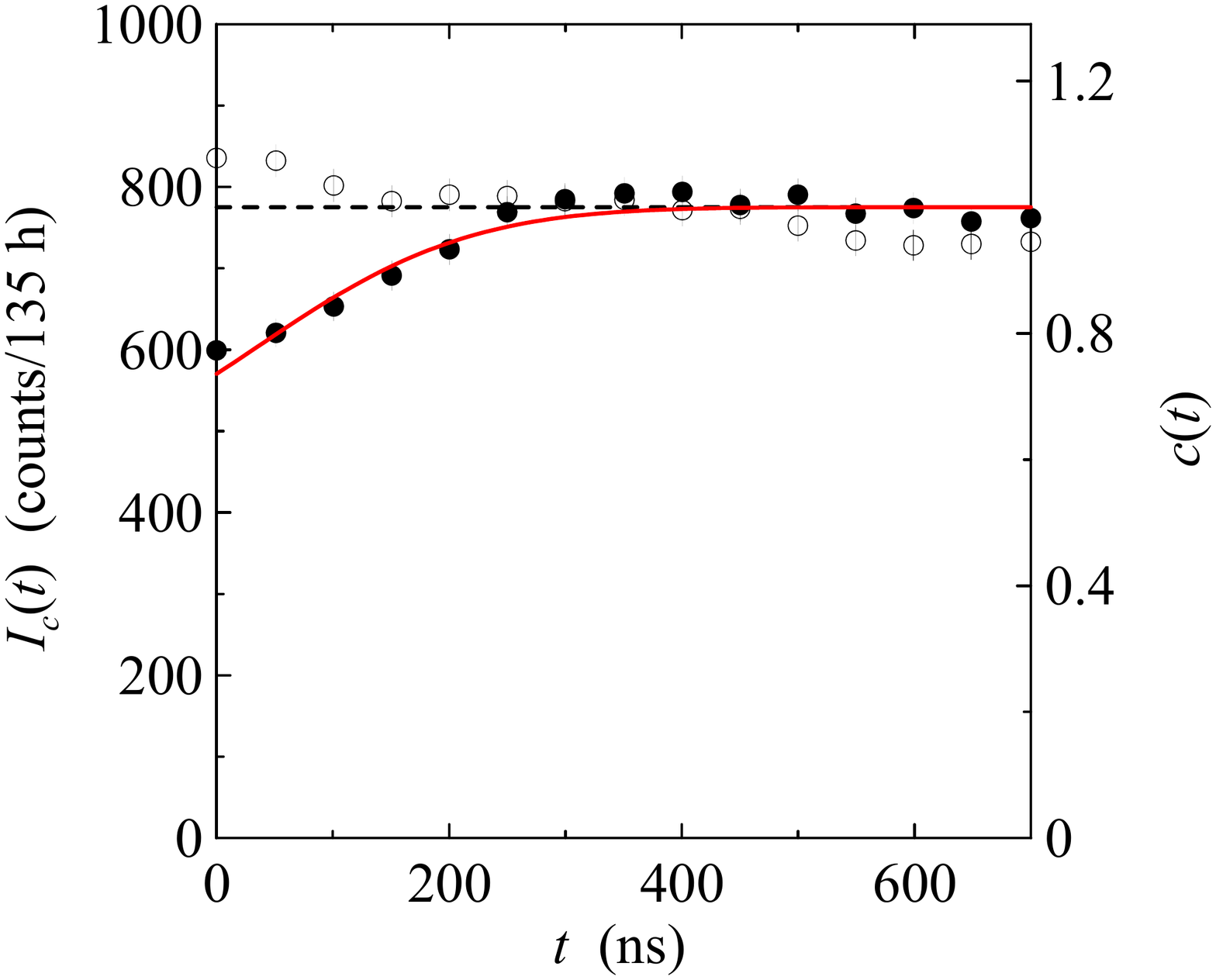}
\end{center}
\vspace{-1.2cm}
\caption{Number of coincidences as a function of the delay time along $z$ at IN10 and T13. One (6-pixel) vertical detector; $\Delta = 400\,$nsec. Symbols and normalization as in Fig.~\ref{fig2}.
}
\label{fig3}
\end{figure}

We report in Fig.\ \ref{fig2}, as a function of $t=t_2-t_1$, the results obtained by applying the above analysis procedure to two pixel groupings equivalent to two vertical detectors 3 $\times$ 0.5 cm$^2$ and 3 cm apart. As we can see, a rather evident minimum is present a $t = 0$. More interesting is the comparison with the results obtained at the T13C beam line where an almost constant response is obtained exactly in the same experimental conditions and by using exactly the same data analysis procedure. In Fig.\ \ref{fig3} we report the results obtained by determining the coincidence rate on a single grouping of pixels of similar size 3 $\times$ 0.5 cm$^2$. In this case the positions where the two neutrons are detected are much less spaced on the average. As it is evident from this plot a clear minimum is again evident and it is more enhanced than in the previous case. The qualitative indication we can derive from the data of Fig.\ \ref{fig3} is that a better effective coherence time is obtained when the two detection points are closer. Again the results obtained on the T13C beam line do not show any minimum in the region close to zero time separation.

To get more quantitative information, assume a Gaussian shape for the functions $1 - c(t) = \alpha \exp(-\beta t^2)$, with $\beta = 1/2 \tau_c^2$, and $W(t) = \sqrt{W / \pi} \exp(-W t^2)$. The coefficient $W$ can be estimated from $W = 1/\bar{\tau}_t^2$, where $\bar{\tau}_t=\sqrt{\langle \tau_t^2 \rangle} \simeq $ 140 ns is the mean square traveling time through the scintillator. From these assumptions we get an analytic expression for the experimental correlation function
\begin{equation}
c_{\mathrm{exp}}(t) = 1 - {\alpha \over 2 \Delta} \, \sqrt{\pi \over \beta} \{\mathrm{erf}[\sqrt{\gamma} (t+ \Delta) ] - \mathrm{erf}(\sqrt{\gamma} t) \},
\label{eq:cexp}
\end{equation}
where the error function $\mathrm{erf}(x)$ is  normalized as usual in such a way that $\mathrm{erf}(0) = 0$ and $\mathrm{erf}(+\infty) = 1$ and $1/\gamma = 1/\beta + 1/W$. We fitted such expression to the experimental data in order to get an estimate for the relevant parameters. As we can see in Figs.\ \ref{fig2} and \ref{fig3}, the fits are fairly good and provide $\tau_c \simeq 30\,$ns and $\tau_c \simeq 120\,$ns for the two estimates of the coherence time. 
The somewhat high value  of the second estimate corresponds to a root mean square energy spread of the incoming beam of the order of 1~neV. In any case, considering the intrinsic approximation we made to derive these estimates, we think that the two results provide the correct order of magnitude and indicate that a transverse coherence effect can be present since the observed zero time minimum is increased when the two detecting points get closer.

In Fig.\ \ref{fig3}, when $t$ is smaller than about 100~ns, the data obtained with short coherence time display a small bump (data taken on T13C). We attribute such small bump to residual cross-talk effects which, in any case, cannot produce the observed minimum in the data obtained at the IN10 beam line with long coherence time.

On the basis of the above results, we can conclude that the present experiment fully confirms the observation of the antibunching effect on a beam of free noninteracting neutrons. It clearly shows, as expected, that the experimental practicability of such an observation strongly depends on the coherence properties of the fermion source: indeed, at the incoherent source T13C the antibunching dip was not detected because the longitudinal coherence time is more than three order of magnitude shorter.

Nonetheless, we still have to face the problem related to the very low density of neutrons in phase space. We believe that the real magnitude of the coherence volume of the beam has to be carefully evaluated, in close relation with the specific experimental conditions under which the measurements are carried out. In particular, the role of the monochromator has to be taken into account, since it strongly modified the shape of the monochromatic neutron wavefunction.

\acknowledgments 
We acknowledge the Institut Laue Langevin of Grenoble for the beam time, that enabled the realization of the experiment. We are grateful to Prof.\ H.\ Rauch for stimulating discussions and helpful suggestions, and to Dr.\ E.\ Santovetti for his invaluable help with the DAQ system. We also thank E.\ Reali for his continuous technical collaboration.


\begin{thebibliography}{99}

\bibitem{bib2} R. Hanbury Brown, R. Q. Twiss, Nature \textbf{177}, 27 (1956).

\bibitem{bib4} W. D\"unnweber, W. Lippich, D. Otten, W. Assmann, K. Hartmann, W. Hering, D. Konnerth, W. Trombik, Phys. Rev. Lett. {\bf 65}, 297 (1990).

\bibitem{bib8} 
S. Y. Kun, R. Gentner, L. Lassen, Z. Phys. A \textbf{342}, 67 (1992);
R. Gentner, K. Keller, W. L{\"u}cking, and L. Lassen, Z. Phys. A \textbf{347}, 401 (1992).

\bibitem{bib5} 
M. Henny, S. Oberholzer, C. Strunk, T. Heinzel, K. Ensslin, M. Holland, and C.
  Sch\"onenberger, Science \textbf{284},  296  (1999).

\bibitem{bib6} 
W.~D. Oliver, J. Kim, R.~C. Liu, and Y. Yamamoto, Science \textbf{284},  299
  (1999).
  
\bibitem{CERN}
F. {Antinori \textit{et~al.} (WA97 Collaboration)}, J. Phys. G \textbf{27},
  2325  (2001);
F. {Antinori \textit{et~al.} (NA57 Collaboration)},
\textit{ibid.} \textbf{34},
  403  (2007).
 
\bibitem{bib7} H. Kiesel, A. Renz and F. Hasselbach, Nature \textbf{418}, 392 (2002). 
  
\bibitem{atom}
T. Rom, {Th. Best}, D. van Oosten, U. Schneider, S. F\"olling, B. Paredes, and
  I. Bloch, Nature (London) \textbf{444},  733  (2006).

\bibitem{bib3} T. Jeltes, J. M. McNamara, W. Hogervorst, W. Vassen, V. Krachmalnicoff, M. Schellekens, A. Perrin, H. Chang, D. Boiron, A. Aspect, C. I. Westbrook, Nature \textbf{445}, 402 (2007), and references therein.

\bibitem{sudarshan}
E.~C.~G. Sudarshan, Phys. Rev. Lett. \textbf{10},  277
  (1963).

\bibitem{glauber}
R.~J. Glauber, Phys. Rev. Lett. \textbf{10},  84  (1963).

\bibitem{bargmann}
V. Bargmann, Commun. Pure Appl. Math. \textbf{14},  187  (1961).

\bibitem{Loudon}
R. Loudon, \textit{The Quantum Theory of Light},
3rd ed. (Oxford University Press, Oxford, 2000).

\bibitem{MandelWolf}
L. Mandel and E. Wolf, \textit{Optical Coherence and Quantum Optics} (Cambridge
  University Press, Cambridge, 1995).

\bibitem{bib13} H. Rauch and S.A. Werner, {\it Neutron Interferometry} (Oxford Science Pubblications, Clarendon Press, 2000).

\bibitem{bib1} M. Iannuzzi, A. Orecchini, F. Sacchetti, P. Facchi, S. Pascazio, Phys. Rev. Lett. \textbf{96} (2006) 080402.

\bibitem{bib9} K. Yuasa \emph{et al},
Phys. Rev. A \textbf{77}, 043623 (2008);
Phys. Rev. B \textbf{79},  180503(R)  (2009);
K. Yuasa, Phys. Rev. B \textbf{80}, 104516 (2009).



\end{thebibliography}
\end{document}